\documentclass[12pt]{iopart}
\usepackage{xcolor}
\usepackage{graphicx} 
\usepackage[utf8]{inputenc}
\usepackage{gensymb}
\begin{document}


\title[Normal coordinates in a system of coupled oscillators] {Normal coordinates in a system of coupled oscillators and  influence of the masses of the springs}

\author{\'Alvaro Su\'arez$^{1}$, Daniel Baccino$^{1}$, Martín Monteiro$^{2}$,
Arturo C. Martí$^3$}

\address{$^1$IPA-ANEP, Montevideo,  Uruguay}
\address{$^2$Universidad ORT Uruguay}
\address{$^3$Instituto de F\'{i}sica,  Universidad de la Rep\'{u}blica, Uruguay}
\ead{alsua@outlook.com}
\ead{dbaccisi@gmail.com}
\ead{mmonteiro@ort.edu.uy}
\ead{marti@fisica.edu.uy}

\date{\today}

\begin{abstract}
Experimental analysis of the motion in a system of two coupled
oscillators with arbitrary initial conditions was performed and the
normal coordinates were obtained directly. The system consisted of two
gliders moving on an air track, joined together by a spring and joined
by two other springs to the fixed ends. From the positions of the
center of mass and the relative distance, acquired by analysis of the
digital video of the experiment, normal coordinates were obtained, and
by a non linear fit the normal frequencies were also obtained. It is
shown that although the masses of the springs are relatively small
compared to that of the gliders, it is necessary to take them into consideration
to improve the  agreement with the experimental results. This experimental-theoretical proposal is targeted to an undergraduate laboratory.
\end{abstract}

\maketitle 

\section{Statement of the problem}
In mechanical systems with several degrees of freedom, in general
terms it is not possible to obtain complete solutions of the equations
of motion. A notable exception to this rule is the dynamics of
conservative systems slightly separated from a point of stable
equilibrium. In this case, the general solutions known as small
oscillations are given by simple periodic solutions or normal modes
characterized by their frequency, known as normal frequency. Knowledge
of these normal modes allows the description in general terms of the
dynamics of the system. Notably, if the initial conditions are
suitably chosen, it is possible to excite the normal modes one by one
and, setting aside perturbations, the system will continue to
oscillate in the same normal mode. This selection of initial
conditions (excepting some systems with symmetries) is not trivial a
priori, and can only be made after having solved the differential
equations of motion. In general terms, given arbitrary initial
conditions, the motion will be a linear combination of all the normal
modes of oscillation. However, there is a method for uncoupling the
equations of motion and expressing them in terms of new variables
called normal coordinates which behave as simple uncoupled
oscillators. The general process for obtaining these normal
coordinates is, from the mathematical point of view, very
straightforward but it often lacks physical meaning. Analysis of
normal modes also plays a very important role in Solid State Physics
where atoms in a crystal lattice are modelled as a set of masses
joined by springs oscillating at preestablished frequencies. The
normal modes of the crystal lattice – phonons – allow explanation of
the electrical and thermodynamic properties of materials.

In university courses, the study of coupled oscillations usually
begins in courses on Waves or Classical Mechanics with systems having
two degrees of freedom. Usually the equations of motion are obtained,
the normal frequencies are calculated and the normal modes of
oscillation are incorporated as a tool to describe the dynamics of the
oscillators \cite{french2001vibrations,crawford1968waves}.  Then the
concept of normal coordinates is introduced as a mathematical
contrivance to uncouple the system of differential equations of motion
and it is shown that they give rise to equations analogous to those of
simple harmonic motion with the frequency of the normal mode. Unlike
normal modes of oscillation, where different experiments are usually
done to excite each particular mode, the use of these normal
coordinates is treated as secondary, without much interest from the
practical point of view.

In more advanced courses of Analytical Mechanics, the study of
oscillating systems is taken up again, but from the perspective of
Lagrange formalism. It is shown in general that for systems with n
degrees of freedom, motion can be described using a set of the same
number of normal coordinates, each oscillating with a defined normal
frequency. Again, as in the basic courses, the absence of experiments
to study specifically the dynamics of normal coordinates could induce
students to hold the false idea that such coordinates are only a
mathematical contrivance.

Several proposals for experiments for analysis of normal modes and
normal coordinates in undergraduate laboratories have been published
\cite{chen1967coupled,wehrbein2001using,greczylo2002using,
givens2003direct,monsoriu2005measuring,castro2013quantitative,singh2019simulating,corridoni2019selective,Tuset_Sanchis_2015,gimenez2017theoretical}.
A pioneering article \cite{chen1967coupled} studied experimentally and
theoretically the arbitrary movements of a system of two gliders in
coupled oscillation. In this work the technological means then
available were used to obtain the changes over time of the coordinates
of each block, after which by solving the equations of motion, the
normal coordinates were obtained and compared with the theoretical
model. In the article by Wehrbein \cite{wehrbein2001using}, which deals
generally with video analysis to study concepts of classical
mechanics, a series of experiments with rigid bodies and oscillating
systems was analysed, among them the monitoring of normal coordinates
for coupled systems with two degrees of freedom. This work used the
means of video analysis available at the time, which only permitted
analysis of slow oscillations, with the center of mass and the
relative position of the oscillators having to be calculated manually
using the coordinates of each mass. Video analysis was also applied later
in the experiment reported in \cite{monsoriu2005measuring} to analyse free and
damped oscillations.
In another proposal, \cite{greczylo2002using}, using an ultrasound sensor
and video analysis tools available at this time, coupled oscillations
in a Wilberforce pendulum were studied. 
It is also worth mentioning
reference \cite{givens2003direct} which proposes a method to observe
each normal mode and to measure the normal frequencies by applying a
variable frequency force in order to attune the system in each normal
mode.

More recently, in a proposal using the sensors in smartphones
\cite{castro2013quantitative,Tuset_Sanchis_2015,gimenez2017theoretical}, 
it is presented a theoretical and experimental study of normal modes in a system of two oscillators, oscillating in a plane, in this case on a table without
friction. From the data acquired by the smartphone sensors the normal
modes and frequencies were obtained and then compared with a
theoretical model.

In the light of this background, it is of interest to propose
laboratory experiments in which the normal coordinates of a system are
obtained directly without perturbing it. This paper proposes an
experiment that permits the normal coordinates to be obtained directly
without perturbing the system, using only the analysis of the videos
of the experiment from which are obtained the time functions, modes
and frequencies. These results compare satisfactorily with those
obtained by theoretical calculation. This experiment can be used in an
advanced undergraduate laboratory class with students who have
knowledge of Classical Mechanics and Oscillations.

\section{Normal modes in coupled systems}
\label{sec:experiment}

We consider a system of two coupled masses joined together and to the
fixed ends with springs as indicated in Figure 1. Our hypothesis is
that the masses of both gliders $M_1$ and $M_2$ are equal to $M$ and
the three elastic constants $k_1$, $k_2$ and $k_3$ are all equal, and
equal to $k$. The gliders can move in the direction indicated and
friction between the gliders and the surface is negligible. We call
the coordinates of each glider with respect to the equilibrium
position $x_1$ and $x_2$, and the distance between the blocks in the
equilibrium position $d$. The equations of motion can be found easily
according to Newton’s second law
\cite{taylor2005classical,morin2008introduction} as:
\begin{equation}
M \frac{d^2 x_1}{dt^2}= - 2 k x_1 +  k x_2,
\label{ecN}
\end{equation}

\begin{equation}
 M \frac{d^2 x_2}{dt^2}=  k x_1 - 2 k x_2. 
 \label{ecN2}
\end{equation}

We can find the normal coordinates directly, without using analytical
mechanics tools, by adding and subtracting the above equations,
obtaining directly an uncoupled system of equations, for the
coordinates $q_S = x_1 + x_2$ and $q_A= x_1 - x_2$:

\begin{equation}
M \frac{d^2 q_S}{dt^2}= - k q_S,
\end{equation}

\begin{equation}
 M \frac{d^2 q_A}{dt^2}=  - 3 k q_A. 
\end{equation}
where $q_S$ and $q_A$ are the normal coordinates of the system, each
having simple harmonic motion with frequencies $w_S= \sqrt{k/M}$ and
$w_A= \sqrt{3k/M}$, independently of the characteristics of the motion
of each of the bodies that make up the system.

We can observe that, as the position of the center of mass measured
from the equilibrium position of glider $1$ is and the position of
glider $2$ with respect to $1$ is $x_{2/1} $, the normal coordinates
$q_S$ and $q_A$ are given by
\begin{equation}
 q_S = 2 x_{cm}- d
 \label{qS}
\end{equation}
\begin{equation}
 q_A =  x_{2/1}- d
 \label{qA}
\end{equation}
So we see that, given the change over time of the center of mass and
the relative position of one glider with respect to the other, we can
easily find the change over time of each normal coordinate, whose
oscillation frequencies $w_S$ and $w_A$ correspond to the frequencies
of the normal modes of symmetric and antisymmetric oscillation in the
system of two gliders, Eqs.~\ref{ecN}-\ref{ecN2}.

\section{Experimental set-up}

The experimental system is composed of two gliders moving on an air
track and three springs, one joining the gliders together and the
other two joining each of the gliders to the ends of the track,
arranged a in linear fashion as shown in Figure 1. The air track
minimizes the friction between the gliders and the track. The air
track and the gliders (SF-9214), the set of 3 springs (ME-9830) and
the air source (SF-9216) were provided by PASCO. The masses of the
gliders, measured on electronic scales, were $M_1=M_2=0.1868(4)$ kg
which within the margin of uncertainty may be considered equal. The
spring constants obtained by static procedures may also be considered
equal within the margin of uncertainty, with $k_1=k_2=k_3=3.02(4)$
N/m.

\begin{figure}
\begin{centering}
\includegraphics[width=0.8\columnwidth]{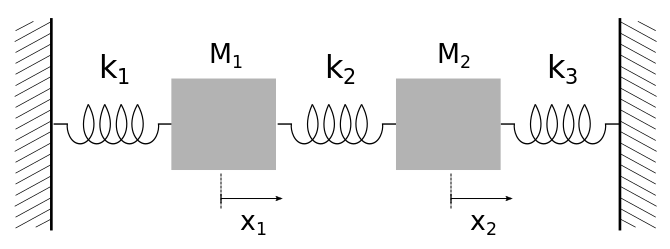}
\par\end{centering}
\caption{\label{figbloques}Diagram of the experimental set-up composed
  of two gliders joined by springs to each other and to the fixed
  ends. To minimize friction the gliders move on an air track.}
\end{figure}

The experiment is as follows: with the power supply of the air track
set to maximum power, the system is moved away from the equilibrium
position and the motion of the gliders is recorded by a digital
camera. In this case the camera built into a Samsung Galaxy S10e
smartphone was used, fixed to a support so that its optical axis was
at a right angle to the track. The digital video obtained was analysed
using Tracker free software \cite{brown2009innovative}. This software
is commonly used to record the motion of point masses in different
situations, for example the bob of a pendulum \cite{MONTEIRO2015}, the
trajectory of a model car \cite{Suarez2019dynamical}, or more
complicated systems \cite{salinas2019dynamics} by recording the
coordinates in the laboratory frame of reference in each video
frame. Other capabilities of the software allow working with a set of
particle systems and obtaining their centers of mass, as well as
studying the relative motions between different particles of the set
\cite{Suarez2019experiment,suarez2020video}. In this work we use these
capabilities to determine the center of mass of a system consisting of
two coupled oscillators, as well as the coordinates of each oscillator
from the point of view of the frame of reference of the other.

\section{Experimental results and discussion}
\begin{figure}
\begin{centering}
\includegraphics[width=0.75\columnwidth]{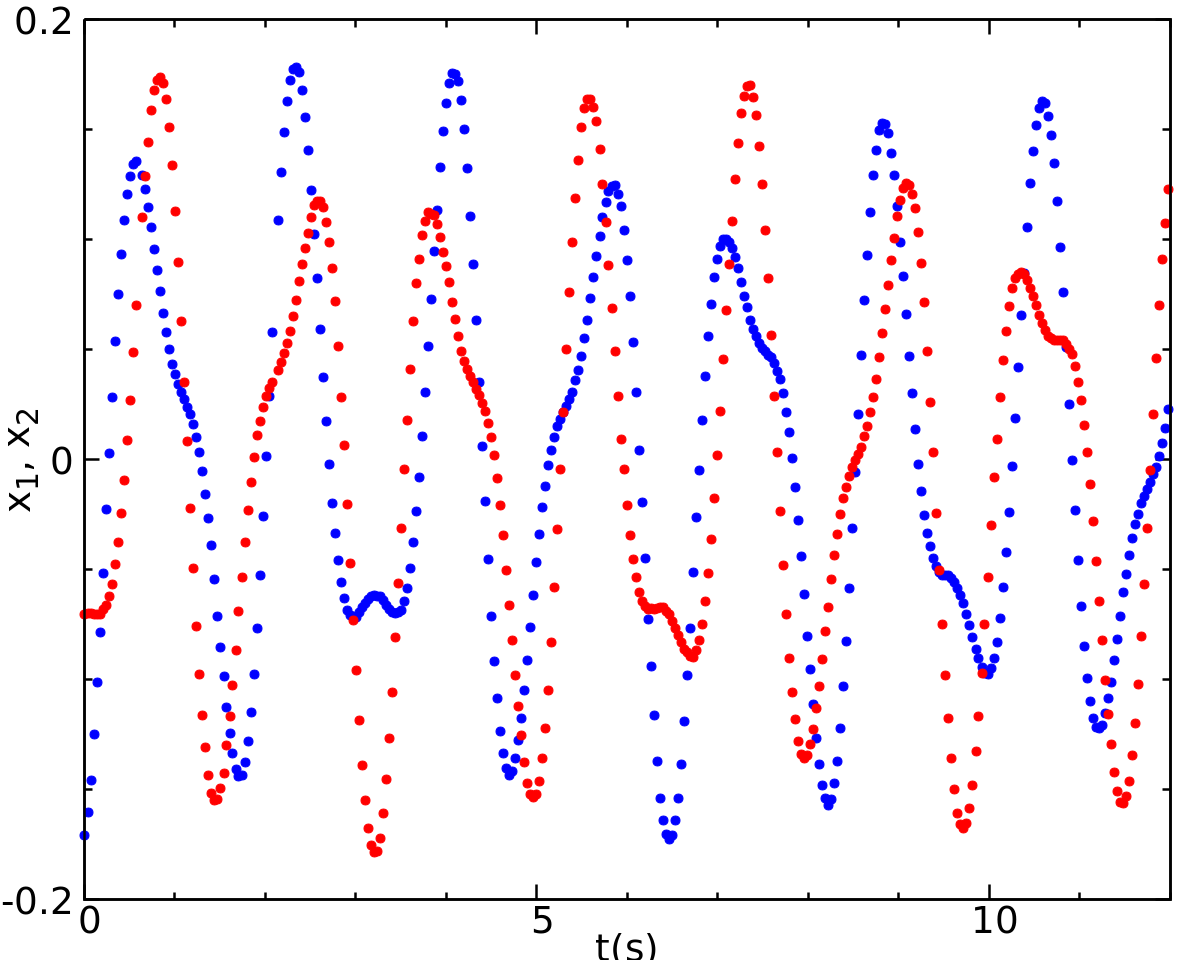}
\par\end{centering}
\caption{\label{figcoor} Temporal evolutions of the
  coordinates of the gliders, $x_1(t) $ (blue) and $x_2(t)$ (red), for
  arbitrary initial conditions.}
\end{figure}

We determined the positions of the gliders, $x_1(t)$ and $x_2(t)$ , by
means of the automatic tracking provided by the Tracker software for
arbitrary initial conditions. Analysis of the video recording of the
experiment gave the changes over time shown in Fig.~\ref{figcoor}. As can be
seen, the motion of the gliders was complex and was not simple
harmonic motion.

\begin{figure}
\begin{centering}
\includegraphics[width=0.78\columnwidth]{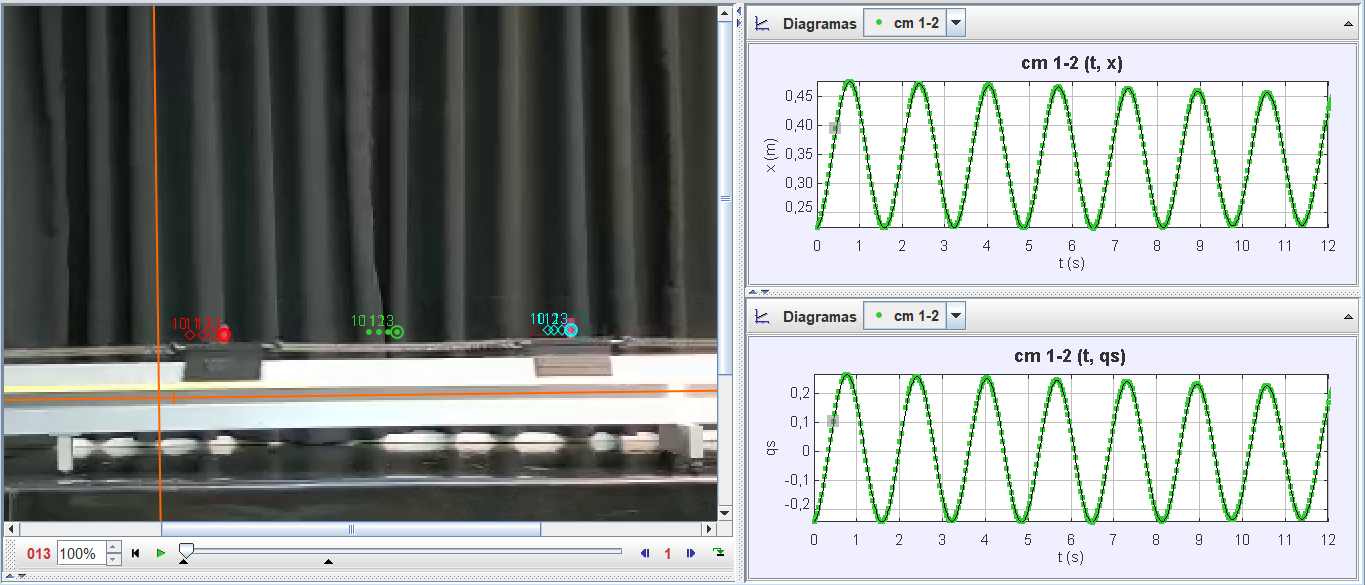}
\includegraphics[width=0.78\columnwidth]{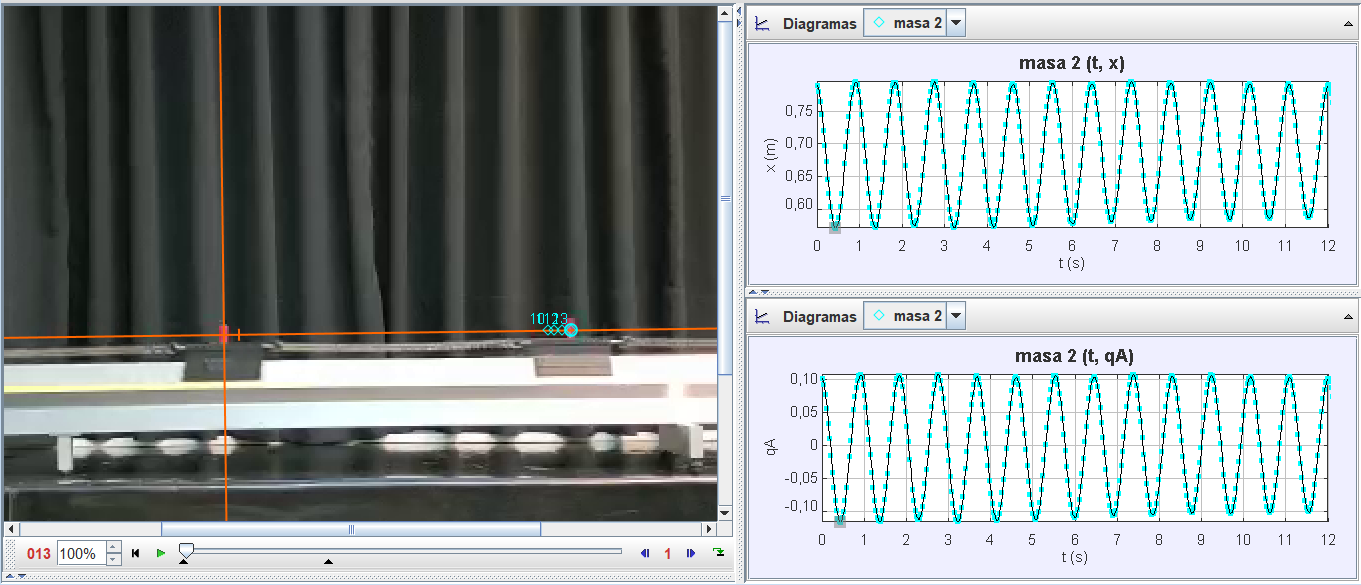}
\par\end{centering}
\caption{\label{figtracker} 
Two Tracker screenshots. The top panel shows the positions of the center of mass and
the symmetric  coordinate $q_S$. The botom panel displays the relative  $x_{2/1}$ and  the antisymmetric $q_A$ coordinates (relative to the glider 1). 
}
\end{figure}

\begin{figure}
\begin{centering}
\includegraphics[width=0.75\columnwidth]{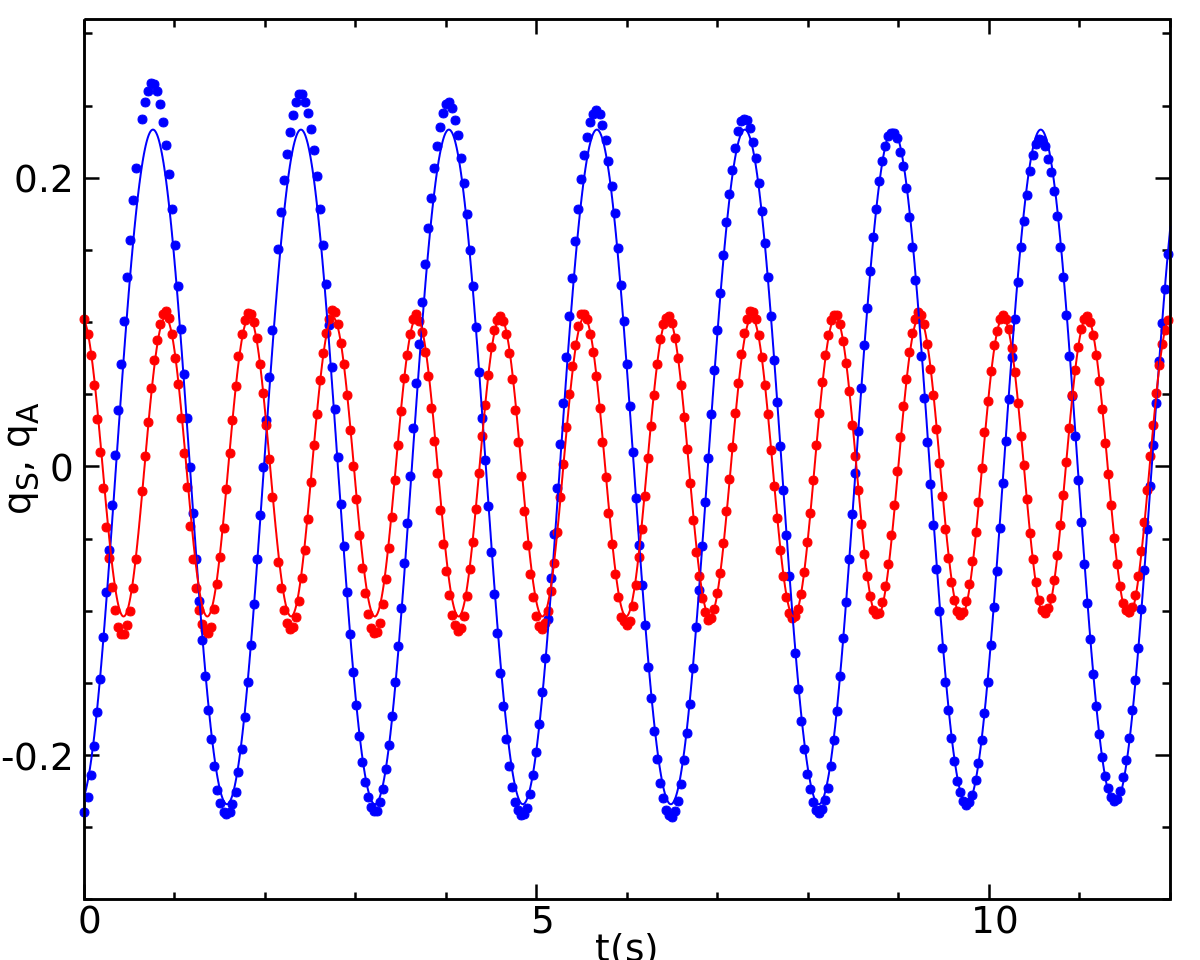}
\par\end{centering}
\caption{\label{figmodosnor} Temporal evolution of the normal
  coordinates $q_S$ (blue circles) and $q_A$ (red circles) and curves
  obtained by non linear fitting to sinusoidal functions (solid
  lines).}\label{figend}
\end{figure}

Afterwards, given the positions of the gliders as a function of time,
using the Tracker software we determined the changes over time of the
position coordinates of the center of mass of the system and of the
motion of glider 2 with respect to glider 1. Figure~\ref{figtracker} displays
two screenshots of the Tracker software. Top panel illustrates the
adquisition of the temporal evolution of the center of mass, $x_{cm}$ and $q_S$,
while the bottom panel that of $x_{2/1}$ and $q_A$. The normal coordinates
$q_S$ and $q_A$ were obtained, thanks to the Tracker, using Eqs.~\ref{qS}-\ref{qA}.
It is worth noting that, as expected, $x_{cm}$ and $x_{2/1}$, proportional
to the $q_S$ and  $q_A$ respectively, follow a sinoudail evolution with the 
corresponding frequencies of the normal coordinates. 
It is remarkable that Tracker allows to fit the sinousoidal curves, and easily obtain the behavior or the normal 
coordinates.  To recapitulate, the temporal evolutions
of the coordinates were determined and fitted to sinusoidal functions, as
shown in Fig.~\ref{figend}. From the parameters of the fitted curves, we obtain the normal frequencies $\omega_S$ and $\omega_A$, with their margin of
uncertainty:
\begin{equation}
 \nonumber
 \omega_S= 3.843(1) rad/s
\end{equation}
\begin{equation}
 \nonumber
 \omega_A=6.789(1) rad/s.
\end{equation}
We will refer to these frequencies as having been obtained by the normal coordinates method.

For a deeper study of the dynamics of the system, we calculated the
normal frequencies by the traditional method which involves setting
the initial conditions so that the gliders oscillate in the normal
modes, first in symmetric and then in antisymmetric motion. We call
this the normal modes method. As with the previous method, by means of
video analysis we obtained the changes over time of the coordinates
(not shown here) and fitted the results to sinusoidal functions. The
frequencies obtained were
\begin{equation}
 \nonumber
 \omega_S= 3.829(1) rad/s
\end{equation}
\begin{equation}
 \nonumber
 \omega_A=6.787(1) rad/s.
\end{equation}
which were slightly different from the values obtained by the normal
coordinates method but always within the uncertainty margin.

\section{Discussion}

In this section we discuss the experimental results and interpret them
in the light of the results of the theoretical model. In the model
presented in section 2, comprising gliders and ideal springs without
mass, the results for the normal frequencies are $\omega_S=4.02(3)$
rad/s and $\omega_A=6.96(5)$ rad/s. These are different from the
experimental measurements by between 2\% and 5\%.  To analyse the
causes of this discrepancy, we discuss below a model that takes into
account the effect of the mass of the springs.

It is not easy to include the effect of the mass of the springs
\cite{chen1957note,fox1970effective,galloni1979influence}. The
pioneering work of Chen \cite{chen1957note} begins with the hypothesis
that the velocity of each spiral in a spring has a linear relationship
with the distance from the fixed end. The Lagrangian of the system is
developed and an overall equation for the normal frequencies of
oscillation is found. In the case of a simple mass-spring system, as a
first approximation the effect of the mass of the spring may be
considered as a disturbance of the mass of the oscillator such that
the effective mass is equal to that of the block plus one-third of the
mass of the spring. This idea can be used to quantify in a simple way
the effect of the mass of the springs on the normal frequencies of
oscillation, when the system oscillates symmetrically and
antisymmetrically, by substituting the system of coupled oscillators
with an equivalent single oscillator having an elastic constant and
effective mass that depend on the parameters of the system.

In the normal mode of symmetric oscillation, the central spring is not
stretched, therefore the blocks of mass $M$ and the spring of mass $m$
behave as a single body of mass $2M+m$.  This new body is joined on
both sides to identical springs, each having elastic constant $k$ and
mass $m$. Thus both springs behave as a single spring with effective
elastic constant $k_{eff}=2k$ and mass $2m$. Using the approximation
that $1/3$ of the mass of each spring is contributed to the total mass
of the system, the effective mass of the equivalent oscillator is
$M_{eff}=2M +m +2m/3$. Then the frequency of symmetric oscillation is:
\begin{equation}
 \nonumber
 \omega_S  
=     \sqrt{\frac{ k }{M +5m/6}}.
\end{equation}

We can carry out a similar analysis for the normal mode of
antisymmetric oscillation. In this case the midpoint of the central
spring is a fixed point, so the system can be divided into two
halves. Each half is composed of a spring with elastic constant $k$
and mass $m$ joined to a block of mass $M$, which in turn is joined to
another spring which is half as long as the original spring, so that
it has elastic constant $2k$ and mass $m/2$. This system then has an
effective elastic constant $k_{eff}=k+2k$ and effective mass
$M_{eff}=M +m/2$. Finally we obtain the frequency of asymmetric
oscillation:
\begin{equation}
 \nonumber
 \omega_A 
  =\sqrt{\frac{ 3k }{M +m/2}}.
\end{equation}

Now we can compare the experimental results obtained by the methods
described above, with the results of the theoretical models which
depend on the spring constants and the masses of the gliders, in the
case which assumes the springs have no mass, as well as the case which
includes a correction for the mass of the springs. The results are
compared in Table 1.

\begin {table}[h]
\label{tab:title} 
\begin{center}
\small
\begin{tabular}{|c|c|c|c|c|} \hline
-- & Normal coordinates & Theoretical model & Theoretical model& Normal modes\\ 
 & method & (massless springs) & (springs with mass) & method\\ \hline
$\omega_S$ & 3.843(1) rad/s & 4.02(3) rad/s & 3.93(3) rad/s & 3.829(1) rad/s\\ \hline
Rel. deviation & -- & 0.046 & 0.022 & 0.0036\\ \hline
$\omega_A$ & 6.789(1) rad/s & 6.96(5) rad/s & 6.87(6) rad/s & 6.787(1) rad/s\\ \hline
Rel. deviation & -- & 0.025 & 0.012 & 0.0003\\ \hline
\end{tabular}
\caption{Comparison between experimental results and models.}
\end{center}
\end {table}

We can see that there is good agreement between the normal frequencies
obtained by all four procedures. The small discrepancy between the
experimental results may be due to the fact that it is not possible to
excite only a single mode and inevitably a mixture occurs with energy
transfer from one mode to the other, alternately. Comparing the
experimental results with the theoretical model there is also very
satisfactory agreement, especially when the correction for the mass of
the springs is taken into account.

\section{Conclusion}
\label{sec:con}
In this work we developed an experimental method which allows direct
visualization of the changes over time of the normal coordinates of a
system comprising two coupled oscillators with arbitrary initial
conditions and studied the influence of the masses of the springs.
Due to the video analysis capabilities of Tracker software
it was easy to find the changes over time of the center of mass of the
system and of the motion of one glider with respect to the other and
afterwards of the normal coordinates. Finally we observed that the
changes of the normal coordinates followed simple harmonic motion and
we measured the frequencies by means of non linear fitting. We
compared the results obtained by exciting each of the normal modes
separately, and compared these with the predictions of theoretical
models with and without correction for the mass of the
springs. Agreement between the experimental results and the
predictions of the theoretical models was very good in all
cases. However, taking into account the mass of the springs improved
considerably  the agreement.

It should be noted that the ease and speed of data processing makes
the activity presented here suitable not only for undergraduate
laboratory courses, but also for direct analysis of the video of the
experiment in theoretical classes, using for example active
methodologies such as Interactive Lecture Demonstrations
\cite{sokoloff1997using} and emphasizing activities with sequences
like POE (Predict – Observe – Explain). Finally, we stress that the
most significant contribution of this work is the possibility of
showing the changes over time of the normal coordinates in advanced
courses of mechanics and waves and allowing students to visualize
their dynamics, showing that these coordinates are not just a mere
mathematical contrivance for solving systems of differential
equations.

 \ack
We acknowledge financial support from grant FSED\_3\_2019\_1\_157320
(ANII-CFE, Uruguay) and CSIC Grupos I+D (UdelaR, Uruguay).

\section*{References} 
\bibliography{/home/arturo/Dropbox/bibtex/mybib}
\bibliographystyle{unsrt}

\end{document}